\documentclass[12pt,a4paper]{article}
\usepackage{latexsym,graphicx,psfrag,amsmath}

\newcommand{\bda}{\begin{\displaymath}\begin{array}{rl}}
\newcommand{\eda}{\end{array}\end{displaymath}}
\newcommand{\be}{\begin{equation}}
\newcommand{\ee}{\end{equation}}
\newcommand{\bdm}{\begin{displaymath}}
\newcommand{\edm}{\end{displaymath}}
\newcommand{\bea}{\begin{eqnarray}}
\newcommand{\eea}{\end{eqnarray}}

\newcommand{\fs}{\,.}
\newcommand{\co}{\,,}

\def\bib#1{\vspace*{-0.2cm}\bibitem{#1}}%
\begin{document}
\begin{center}
{\LARGE \bf On the history of the strong interaction} \\

\vspace{0.8cm}
H.~Leutwyler\\ Albert Einstein Center for Fundamental Physics\\Institute for Theoretical Physics, University of Bern\\
Sidlerstr.~5, CH-3012 Bern, Switzerland
\end{center}
\vspace{0.2cm}

\begin{abstract}These lecture notes recall the conceptual developments which led from the discovery of the neutron to our present understanding of strong interaction physics.  \end{abstract} 

\begin{center}
Lectures given at the International School of Subnuclear Physics\\  Erice, Italy, 23 June -- 2 July 2012
\end{center}

\footnotesize
\tableofcontents
\normalsize
\section{From nucleons to quarks}
I am not a historian. The following text describes my own recollections and is mainly based on memory, which unfortunately does not appear to improve with age \ldots\, For a professional account, I refer to the book by Tian Yu Cao \cite{Cao}.  A few other sources are referred to below.

\subsection{Beginnings}\label{beginnings}
The discovery of the neutron by Chadwick in 1932 may be viewed as the birth of the strong interaction: it indicated that the nuclei consist of protons and neutrons and hence the presence of a force that holds them together, strong enough to counteract the electromagnetic repulsion. Immediately thereafter, Heisenberg introduced the notion of isospin as a symmetry of the strong interaction, in order to explain why proton and neutron nearly have the same mass \cite{Heisenberg}. In 1935, Yukawa pointed out that the nuclear force could be generated by the exchange of a hypothetical spinless particle, provided its mass is intermediate between the masses of proton and electron -- a {\it meson} \cite{Yukawa}.  Today, we know that such a particle indeed exists: Yukawa predicted the pion. Stueckelberg pursued similar ideas, but was mainly thinking about particles of spin 1, in analogy with the particle that mediates the electromagnetic interaction \cite{Stueckelberg}.   

In the thirties and fourties of the last century, the understanding of the force between two nucleons made considerable progress, in the framework of nonrelativistic potential models. These are much more flexible than quantum field theories. Suitable potentials that are attractive at large distances but repulsive at short distances do yield a decent understanding of nuclear structure: Paris potential, Bonn potential, shell model of the nucleus. In this framework, nuclear reactions, in particular the processes responsible for the luminosity of the sun, stellar structure, $\alpha$-decay and related matters were well understood more than fifty years ago. 

These phenomena concern interactions among nucleons with small relative velocities. Experimentally, it had become possible to explore relativistic collisions, but a description in terms of nonrelativistic potentials cannot cover these. In the period between 1935 and 1965, many attempts at formulating a theory of the strong interaction based on elementary fields for baryons and mesons were made. In particular, uncountable PhD theses were written, based on local interactions of the Yukawa type, using perturbation theory to analyze them, despite the fact that the coupling constants invariably turned out to be numerically large. Absolutely nothing worked even half way. 

Although there was considerable progress in understanding the general principles of quantum field theory (Lorentz invariance, unitarity, crossing symmetry, causality, analyticity, dispersion relations, CPT theorem, spin and statistics) as well as in renormalization theory, faith in quantum field theory was in decline, even concerning QED (Landau-pole). To many, the renormalization procedure needed to arrive at physically meaningful results looked suspicious and it appeared doubtful that the strong interaction could at all be described by means of a local quantum field theory.  Some suggested that this framework should be replaced by S-matrix theory -- heated debates concerning this suggestion took place at the time \cite{Pietschmann}. Regge poles were considered as a promising alternative to the quantum fields (the Veneziano model is born in 1968 \cite{Veneziano}). Fifty years ago, when I completed my studies, the quantum field theory of the strong interaction consisted of a collection of beliefs, prejudices and assumptions. Quite a few of these turned out to be wrong.

\subsection{Flavour symmetries}
Symmetries that extend isospin to a larger Lie group provided the first hints towards an understanding of the structure underneath the strong interaction phenomena. The introduction of the strangeness quantum number and the Gell-Mann-Nishijima formula \cite{Gell-Mann Nishijima} was a significant step in this direction. Goldberger and Treiman \cite{Goldberger Treiman} then showed that the {\it axial-vector current} plays an important role, not only in the weak interaction (the  pion-to-vacuum matrix element of this current -- the pion decay constant $F_\pi$ -- determines the rate of the weak decay $\pi\rightarrow\mu\nu$) but also in the context of the strong interaction: the nucleon matrix element of the axial-vector current, $g_A$, determines the strength of the interaction between pions and nucleons: $g_{\pi N}=g_A M_N/F_\pi$. 

At low energies, the main characteristic of the strong interaction is that the energy gap is small: the lightest state occurring in the eigenvalue spectrum of the Hamiltonian is the pion, with $M_\pi\simeq $ 135 MeV, small compared to the mass of the proton, $M_p\simeq$ 938 MeV. In 1960, Nambu found out why that is so: it has to do with a hidden, approximate, continuous symmetry. Since some of its generators carry negative parity, it is referred to as a {\it chiral symmetry}. For this symmetry to be consistent with observation, it is essential that an analog of spontaneous magnetization occurs in particle physics: for dynamical reasons, the  state of lowest energy -- the vacuum -- is not symmetric under chiral transformations. Consequently, the symmetry cannot be seen in the spectrum of the theory: it is {\it hidden} or {\it spontaneously broken}. Nambu realized that the spontaneous breakdown of a continuous symmetry entails massless particles analogous to the spin waves of a magnet and concluded that the pions must play this role. If the strong interaction was strictly invariant under chiral symmetry,  there would be no energy gap at all -- the pions would be massless.\footnote{A precise formulation of this statement, known as the {\it Goldstone theorem}, was given later \cite{Goldstone theorem}.}  Conversely, since the pions are not massless, chiral symmetry cannot be exact -- unlike isospin, which at that time was taken to be an exact symmetry of the strong interaction. The spectrum does have an energy gap because chiral symmetry is not exact: the pions are not massless, only light. In fact, they represent the lightest strongly interacting particles that can be exchanged between two nucleons. This is why, at large distances, the potential between two nucleons is correctly described by the Yukawa formula.  

The discovery of the {\it Eightfold Way} by Gell-Mann and Ne'eman paved the way to an understanding of the mass pattern of the baryons and mesons \cite{Ne'eman,Gell-Mann SU3}. Like chiral symmetry, the group SU(3) that underlies the Eightfold way represents an approximate symmetry: the spectrum of the mesons and baryons does not consist of degenerate multiplets of this group. The splitting between the energy levels, however, does exhibit a pattern that can be understood in terms of the assumption that the part of the Hamiltonian that breaks the symmetry transforms in a simple way. This led to the Gell-Mann-Okubo formula \cite{Gell-Mann SU3,Okubo} and to a prediction for the mass of the $\Omega^-$, a member of the baryon decuplet which was still missing, but was soon confirmed experimentally, at the predicted place \cite{Omega minus}.  

\subsection{Quark Model}
In 1964, Gell-Mann \cite{Gell-Mann Quarks} and Zweig \cite{Zweig} pointed out that the observed pattern of baryons can qualitatively be understood on the basis of the assumption that these particles are bound states built with three constituents, while the spectrum of the mesons indicates that they contain only two of these. Zweig called the constituents ``aces''. Gell-Mann coined the term ``quarks'', which is now commonly accepted. The Quark Model gradually evolved into a very simple and successful semi-quantitative framework, but gave rise to a fundamental puzzle: why do the constituents not show up in experiment ? For this reason, the existence of the quarks was considered doubtful: ``Such particles [quarks] presumably are not real but we may use them in our field theory anyway \ldots '' \cite{Gell-Mann Physics}.  Quarks were treated like the veal used to prepare a pheasant in the royal french cuisine: the pheasant was baked between two slices of veal, which were then discarded (or left for the less royal members of the court). Conceptually, this was a shaky cuisine.

If the flavour symmetries are important, why are they not exact ? Gell-Mann found a beautiful explanation: {\it current algebra} \cite{Gell-Mann SU3,Gell-Mann Physics}. The charges form an exact algebra even if they do not commute with the Hamiltonian and the framework can be extended to the corresponding currents, irrespective of whether or not they are conserved. Adler and Weisberger showed that current algebra can be tested with the sum rule that follows from the nucleon matrix element of the commutator of two axial-vector charges \cite{Adler Weisberger}. Weinberg then demonstrated that even the strength of the interaction among the pions can be understood on the basis of current algebra: the $\pi\pi$ scattering lengths can be predicted in terms of the pion decay constant \cite{Weinberg scattering lengths}.  

\subsection{Behaviour at short distances}\label{short distances}
Bjorken had pointed out that if the nucleons contain point-like constituents, then the $ep$ cross section should obey scaling laws in the deep inelastic region \cite{Bjorken}. Indeed, the scattering experiments carried out by the MIT-SLAC collaboration in 1968/69 did show experimental evidence for such constituents \cite{Friedman Kendall Taylor}. Feynman called these {\it partons}, leaving it open whether they were the quarks or something else.

The operator product expansion turned out to be a very useful tool for the short distance analysis of the theory -- the title of the paper where it was introduced \cite{Wilson}, ``Nonlagrangian models of current algebra'', reflects the general skepticism towards Lagrangian quantum field theory that I mentioned at the end of section \ref{beginnings}.

\subsection{Colour}
The Quark Model was difficult to reconcile with the spin-statistics theorem which implies that particles of spin $\frac{1}{2}$ must obey Fermi statistics. Greenberg proposed that the quarks obey neither Fermi-statistics nor Bose-statistics, but ``para-statistics of order three'' \cite{Greenberg}. The proposal amounts to the introduction of a new internal quantum number. Indeed, in 1965, Bogoliubov, Struminsky and Tavkhelidze \cite{BST}, Han and Nambu \cite{HN} and Miyamoto \cite{Miyamoto} independently pointed out that some of the problems encountered in the quark model disappear if the $u$, $d$ and $s$ quarks occur in 3 states. Gell-Mann coined the term ``colour'' for the new quantum number. 

One of the possibilities considered for the interaction that binds the quarks together was an abelian gauge field analogous to the e.m.~field, but this gave rise to problems, because the field would then interfere with the other degrees of freedom. Fritzsch and Gell-Mann pointed out that if the gluons carry colour, then the empirical observation that quarks appear to be confined might also apply to them: the spectrum of the theory might exclusively contain colour neutral states. 

In his lectures at the Schladming Winter School in 1972 \cite{Gell-Mann Schladming}, Gell-Mann thoroughly discussed the role of the quarks and gluons: theorists had to navigate between Scylla and Charybdis, trying to abstract neither too much nor too little from models built with these objects. The basic tool at that time was  {\it current algebra on the light cone}. 
He invited me to visit Caltech. I did that during three months in the spring break of 1973 and spent an extremely interesting period there.  
 
\subsection{QCD}
Gell-Mann's talk at the High Energy Physics Conference in 1972 (Fermilab), had the title ``Current algebra: Quarks and what else?'' In particular, he discussed the proposal to describe the gluons in terms of a {\it nonabelian gauge field} coupled to colour, relying on work done with Fritzsch \cite{Gell-Mann Fermilab}.  As it was known already that the electromagnetic and weak interactions are mediated by gauge fields, the idea that colour might be a local symmetry as well does not appear as far fetched. In the proceedings, Fritzsch and Gell-Mann mention unpublished work in this direction by Wess. For a recent outline of the development that led from the quark model to {\it Quantum Chromodynamics} -- the quantum field theory which describes the interaction among the quarks and gluons -- see \cite{Ecker,Fritzsch}. 

The main problem at the time was that for a gauge field theory to describe the hadrons and their interaction, it had to be fundamentally different from the quantum field theories encountered in nature so far: all of these, including the electroweak theory, have the spectrum indicated by the degrees of freedom occurring in the Lagrangian: photons, leptons, intermediate bosons, \ldots\,  The proposal can only make sense if this need not be so, that is if the spectrum of physical states in a quantum field theory can differ from the spectrum of the fields needed to formulate it: gluons and quarks in the Lagrangian, hadrons in the spectrum. This looked like wishful thinking. 
\section{On the history of the gauge field concept}
\subsection{Electromagnetic interaction} 
The final form of the laws obeyed by the electromagnetic field was found by Maxwell, around 1860 -- these laws
survived relativity and quantum theory, unharmed.

Fock pointed out that the Schr\"odinger equation for electrons in an electromagnetic field, 
\bdm
\frac{1}{i}\,\frac{\partial\psi}{\partial t}-\frac{1}{2 m_e^2}\,
(\vec{\nabla}
+i\, e\vec{A})^2\psi- e\, \varphi\, \psi=0\co\edm
is invariant under a group of local transformations:
\bdm
\vec{A}^{\,\prime}(x)=\vec{A}(x)+\vec{\nabla}\alpha(x),\hspace{0.5cm}
\displaystyle\varphi^{\,\prime}(x)=
\varphi(x)-\frac{\partial \alpha(x)}{\partial t},\hspace{0.5cm}
 \psi(x)'=e^{- i e \alpha(x)}\,\psi(x)\co\edm
in the sense that the fields $\vec{A}^\prime,\varphi^\prime,\psi^\prime$ describe the same physical situation as
$\vec{A},\varphi,\psi$ \cite{Fock}. Weyl termed these  {\it gauge transformations} (with gauge group U(1) in this case).
In fact, the electromagnetic interaction is fully characterized by the symmetry with respect to this group: gauge invariance is the crucial property of this interaction. 

I illustrate the statement with the core of Quantum Electrodynamics: photons and electrons. 
Gauge invariance allows only two free parameters in the Lagrangian of this system: $e,m_e$. 
Moreover, only one of these is dimensionless: $e^2/4\pi=1/137.035\, 999\, 074\, (44)$.  U(1) symmetry and renormalizability fully determine the properties of the e.m.~interaction, except for this number, which so far still remains unexplained.\footnote{As a side remark, I note that in QED, an additional term, proportional to $F_{\mu\nu}\tilde{F}^{\mu\nu}$ can be added to the Lagrangian.  
The term represents a total derivative,  $F_{\mu\nu}\tilde{F}^{\mu\nu}=\partial_\mu f^\mu$.
Since e.m. fields with a nontrivial behaviour at large distances do not appear to play a significant role (no instantons or the like), this term does not affect the physics.}

\subsection{Gauge fields from geometry}
 
 Kaluza \cite{Kaluza} and Klein \cite{Klein} had shown that a 5-dimensional Riemann space with a metric that is independent of the fifth coordinate is equivalent to a 4-dimensional world with {\it gravity}, a {\it gauge field} and a {\it scalar field}. In this framework, gauge transformations amount to a shift in the fifth direction:
$x^{5'}=x^5+ \alpha(\vec{x},t)$. In geometric terms, a metric space of this type is characterized by a group of  isometries: the geometry remains the same along certain directions, indicated by Killing vectors. In the case of the 5-dimensional spaces considered by Kaluza and Klein, the isometry group is the abelian group U(1). The fifth dimension can be compactified to a circle -- U(1) then generates motions on this circle. A particularly attractive feature of this theory is that it can explain the quantization of the electric charge: fields living on such a manifold necessarily carry integer multiples of a basic charge unit. 

Pauli noticed that the Kaluza-Klein scenario admits a natural generalization to higher dimensions, where larger isometry groups find place. Riemann spaces of dimension $> 5$ admit nonabelian isometry groups that reduce the system to a 4-dimensional one with {\it gravity}, {\it nonabelian gauge fields} and several {\it scalar fields}. 
Pauli was motivated by the isospin symmetry of the meson-nucleon interaction and focused attention on a Riemann space of dimension 6, with isometry group SU(2). 

Pauli did not publish the idea that the strong interaction might arise in this way, because he was convinced that the quanta of a gauge field are massless: gauge invariance does not allow one to put a mass term into the Lagrangian. He concluded that the forces mediated by gauge fields are necessarily of long range and can therefore not mediate the strong interaction, which is known to be of short range. More details concerning Pauli's thoughts can be found in \cite{Straumann}.

\subsection{Nonabelian gauge fields}
The paper of Yang and Mills appeared in 1954 \cite{Yang and Mills}. Ronald Shaw, a student of Salam, independently formulated nonabelian gauge field theory in his PhD thesis \cite{Shaw}. Ten years later,
Higgs \cite{Higgs}, Brout and  Englert \cite{Brout and Englert} and Guralnik, Hagen and Kibble \cite{Guralnik et al} showed that Pauli's objection is not valid in general: in the presence of scalar fields, gauge fields can pick up mass, so that forces mediated by gauge fields can be of short range. The work of Glashow \cite{Glashow}, Weinberg \cite{Weinberg} and Salam \cite{Salam} then demonstrated that nonabelian gauge fields are relevant for physics: the framework discovered by Higgs et al. does accommodate a realistic description of the e.m.\,and weak interactions.
 
\subsection{Asymptotic freedom}
  Already in 1965, Vanyashin and Terentyev \cite{Vanyashin and Terentyev} found that the renormalization of the electric charge of a vector field is of opposite sign to the one of the electron. In the language of SU(2) gauge field theory, their result implies that the $\beta$-function is negative at one loop. 

The first correct calculation of the $\beta$-function of a nonabelian gauge field theory was carried out by Khriplovich, for the case of SU(2), relevant for the electroweak interaction \cite{Khriplovich}. He found that $\beta$ is negative and concluded that the interaction becomes weak at short distance. In his PhD thesis, 't Hooft performed the calculation of the $\beta$-function for an arbitrary gauge group, including the interaction with fermions and Higgs scalars \cite{'t Hooft}. He demonstrated that the theory is renormalizable and confirmed that, unless there are too many fermions or scalars, the $\beta$-function is negative at small coupling.  

%\subsection{Implications for QCD}
In 1973, Gross and Wilczek \cite{Gross and Wilczek} and Politzer \cite{Politzer} discussed the consequences of a negative $\beta$-function and suggested that this might explain Bjorken scaling, which had been observed at SLAC in 1969 (cf.~section \ref{short distances}). They pointed out that QCD predicts specific modifications of the scaling laws. In the meantime, there is strong experimental evidence for these.  

A detailed account of the history of the quantum theory of gauge fields can be found in the 1998 Erice lectures of 't Hooft \cite{tHooft Erice}. 

\section{Quantum Chromodynamics}
\subsection{Arguments in favour of QCD}
Reasons for proposing QCD as a theory of the strong interaction were given in \cite{Fritzsch Gell-Mann Leutwyler}.
The idea that the observed spectrum of particles can fully be understood on the basis of a theory built with quarks and gluons still looked rather questionable and was accordingly formulated in cautious terms. In the abstract, for instance, it is pointed out that "\ldots there are several advantages in abstracting properties of hadrons and their currents from a Yang-Mills gauge model based on coloured quarks and colour octet gluons." Before the paper was completed, preprints by Gross and Wilczek \cite{Gross and Wilczek} and Politzer \cite{Politzer} were circulated - they are quoted and asymptotic freedom is given as argument $\#$4 in favour of QCD. Also, important open questions were pointed out, in particular, the U(1) problem. 
 
 Many considered QCD a wild speculation. On the other hand, several papers concerning gauge field theories that include the strong interaction appeared around the same time, for instance \cite{Pati and Salam,Weinberg QCD}.
\subsection{November revolution}
The discovery of the $J/\psi$ was announced simultaneously at Brookhaven and SLAC, on November 11, 1974. 
Three days later, the observation was confirmed at ADONE, Frascati and ten days later, the $\psi'$ was found at SLAC, where subsequently many further related states were discovered. We now know that these are bound states formed with the $c$-quark and its antiparticle and that there are two further heavy quarks: $b$, $t$. 

At sufficiently high energies, quarks and gluons do manifest themselves as jets. Like the neutrini, they have left their theoretical place of birth and can now be seen flying around like ordinary, observable particles.  
Gradually, particle physicists abandoned their outposts in no man's and no woman's land, returned to the quantum fields and resumed discussion in the good old {\it Gasthaus zu Lagrange}, a term coined by Jost. The theoretical framework that describes the strong, electromagnetic and weak interactions in terms of gauge fields, leptons and quarks is now referred to as the Standard Model - this framework clarified the picture enormously.\footnote{Indeed, the success of this theory is amazing: Gauge fields are renormalizable in four dimensions, but it looks unlikely that the Standard Model is valid much beyond the explored energy range. Presumably it represents an effective theory. There is no reason, however, for an effective theory to be renormalizable. One of the most puzzling aspects of the Standard Model is that it is able to account for such a broad range of phenomena that are characterized by very different scales within one and the same renormalizable theory.}

\subsection{Theoretical paradise}\label{theoretical paradise}
In order to briefly discuss some of the basic properties of QCD, let me turn the electroweak interaction off, treat the three light quarks as massless and the remaining ones as infinitely heavy:
$$m_u=m_d=m_s=0\co\hspace{1cm}m_c=m_b=m_t=\infty\fs$$ 
The Lagrangian then contains a single parameter: the coupling constant $g$, which may be viewed as the net colour of a quark. Unlike an electron, a quark cannot be isolated from the rest of the world -- its colour 
$g$ depends on the radius of the region considered. According to perturbation theory, the colour contained in a sphere of radius $r$ grows logarithmically with the radius:\footnote{The formula only holds if the radius is small, $r \,\Lambda_{QCD}\ll1$.} 
$$\alpha_s\equiv\frac{g^2}{4\pi} =\frac{2\pi}{9\,|\ln (r\,\Lambda_{QCD})|}\fs$$
Although the classical Lagrangian of  massless QCD does not contain any dimensionful parameter, the corresponding quantum field theory does: the strength of the interaction cannot be characterized by a number, but by a dimensionful quantity, the intrinsic scale $\Lambda_{QCD}$. 

The phenomenon is referred to as {\it dimensional transmutation}. In perturbation theory, it manifests itself through the occurrence of divergences -- contrary to what many quantum field theorists thought for many years, the divergences do not represent a disease, but are intimately connected with the structure of the theory. They are a consequence of the fact that a quantum field theory does not inherit all of the properties of the corresponding classical field theory. In the case of massless Chromodynamics, the classical version is conformally invariant, but the quantum version is not. This is crucial for QCD to be consistent with what is known about the strong interaction. In fact, massless QCD is how theories should be: it does not contain a single dimensionless parameter. In principle, the values of all quantities of physical interest are predicted without the need to tune parameters (the numerical value of the mass of the proton in kilogram units cannot be calculated, of course, because that number depends on what is meant by a kilogram, but the mass spectrum, the width of the resonances, the cross sections, the form factors, \ldots\, can be calculated in a parameter free manner from the mass of the proton, at least in principle). 
 
\subsection{Symmetries of massless QCD}
The couplings of the  $u$-, $d$- and $s$-quarks to the gauge field are identical.
If the masses are set equal to zero, there is no difference at all -- the Lagrangian is symmetric under SU(3) rotations in flavour space.  Indeed, there is more symmetry: for massless fermions, the right- and left-handed components can be subject to independent flavour rotations. The Lagrangian of QCD with three massless flavours is invariant under SU(3)$_{R}\times$SU(3)$_{L} $. QCD thus explains the presence of the mysterious chiral symmetry discovered by Nambu: an exact symmetry of this type is present if some of the quarks are massless. 

Nambu had conjectured that chiral symmetry breaks down spontaneously. Can it be demonstrated that   the symmetry group SU(3)$_{R}$$\times$SU(3)$_{L}$ of the Lagrangian of massless QCD spontaneously break down to the subgroup SU(3)$_{R+L}$ ? To my knowledge an analytic proof is not available, but the work done on the lattice demonstrates beyond any doubt that this does happen. In particular, for $m_u=m_d=m_s$, the states do form degenerate multiplets of SU(3)$_{R+L}$ and, in the limit $m_u,m_d,m_s\rightarrow 0$, the pseudoscalar octet does become massless, as required by the Goldstone theorem.

\subsection{Quark masses}
The 8 lightest mesons,  $\pi^+,\pi^0,\pi^-,K^+,K^0,\bar{K}^0,K^-,\eta$, do have the quantum numbers of the Nambu-Goldstone bosons, but massless they are not. The reason is that we are not living in the paradise described above: the light quark masses are different from zero. Accordingly, the Lagrangian of QCD is only approximately invariant under chiral rotations, to the extent that the symmetry breaking parameters $m_u\,,\,m_d\,,\, m_s$ are small. Hence the multiplets are split; in particular, the Nambu-Goldstone boson multiplet is not massless.

% \subsection{Pattern of light quark masses}

 Even before the discovery of QCD, attempts at estimating the masses of the quarks were made. In particular, nonrelativistic bound state models for mesons and baryons where constructed. In these models, the proton mass is dominated by the sum of the masses of its constituents:
$m_u +m_u + m_d\simeq M_p$, $m_u\simeq m_d\simeq 300 \,\mbox{MeV}$.  
With the discovery of QCD, the mass of the quarks became an unambiguous  concept: the quark masses occur in the Lagrangian of the theory.  The first crude estimate within QCD relied on a model for the wave functions of $\pi$, $K$, $\rho$, which was based on SU(6) (spin-flavour-symmetry) and led to \cite{5 MeV}
$${\displaystyle \frac{(m_u+m_d)}{2}= \frac{F_\pi
  M_\pi^2}{3 F_\rho M_\rho}\simeq 5\,\mbox{MeV},\rule{1.1cm}{0cm}
m_s \simeq 135\,\mbox{MeV}}\,.\hspace{1.5cm} $$
Similar mass patterns were found earlier, within the Nambu-Jona-Lasinio model \cite{NJL} or on the basis of sum rules \cite{Okubo 1969}.

From the time Heisenberg had introduced isospin symmetry, it was taken for granted that the strong interaction strictly conserves isospin. QCD does have this symmetry if and only if $m_u=m_d$. If that condition were met, the mass difference between the proton and the neutron would be due exclusively to the e.m.\,interaction. When J\"urg Gasser and I analyzed this mass difference by means of the Cottingham formula, however, we came to the conclusion that this cannot be so: evaluating the formula on the basis of Bjorken scaling, we found that the electromagnetic self energy of the proton is larger than the one of the neutron. In fact, a very strong breaking of isospin symmetry in the quark masses is needed to explain the observed value of $M_n-M_p$ within QCD \cite{GL 1975}: 
$$m_u\simeq 4 \,\mbox{MeV},\hspace{0.5em}m_d\simeq 7
  \,\mbox{MeV},\hspace{0.5em}m_s\simeq 135\,\mbox{MeV}\,.\hspace{1.5cm}$$
In other words, QCD can be consistent with experiment only if $m_u$ and $m_d$ are very different. It took quite a while before this bizarre pattern was generally accepted. The Dashen theorem \cite{Dashen} states that, in a world where the quarks are massless, the e.m.\,self-energies of the kaons and pions obey the relation $M_{K^+}^{2\;em}-M_{K^0}^{2\;em}=M_{\pi^+}^{2\;em}-M_{\pi^0}^{2\;em}\!.$ If the mass differences were dominated by the e.m.\,interaction, the charged kaon would be heavier than the neutral one. Hence the mass difference between the kaons cannot be due to the electromagnetic interaction, either. The estimates for the quark mass ratios obtained with the Dashen theorem confirm the above pattern \cite{Weinberg 1977}.
 
\subsection{Approximate symmetries are natural in QCD}
At first sight, the fact that $m_u$ strongly differs from $m_d$ is puzzling: if this is so, why is isospin such a good quantum number ? The key observation here is the one discussed in section \ref{theoretical paradise}: QCD has an intrinsic scale, $\Lambda_{QCD}$. For isospin to represent an approximate symmetry, it is not necessary that $m_d-m_u$ is small compared to $m_u+m_d$. It suffices that the symmetry breaking parameter  is small compared to the intrinsic scale, $m_d-m_u\ll \Lambda_{QCD}$.
 
In the case of the eightfold way, the symmetry breaking parameters are the differences between the masses of the three light quarks. If they are small compared to the intrinsic scale of QCD, then the Green functions, masses, form factors, cross sections \ldots\,are approximately invariant under the group SU(3)$_{R+L}$. Isospin is an even better symmetry, because the relevant symmetry breaking parameter is smaller, $m_d-m_u\ll m_s-m_u$. The fact that $M_{\pi^+}^2$ is small compared to $M_{K^+}^2$ implies $m_u+m_d\ll m_u+m_s$. Hence all three light quark masses must be small compared to the scale of QCD. 

In the framework of QCD, the presence of an approximate chiral symmetry group of the form SU(3)$_R\times$SU(3)$_L$ thus has a very simple explanation: it so happens that the masses of $u$, $d$ and $s$ are small. We do not know why, but there is not doubt that this is so. The quark masses represent a perturbation, which in first approximation can be neglected -- in first approximation, the world is the paradise described above.
 
\section{ Conclusion}
We now know the origin of the strong interaction. The gauge field theory that describes it, QCD, represents the first non-trivial quantum field theory that is internally consistent in four space-time dimensions. In my opinion, this constitutes the most beautiful part of the Standard Model.  Apart from the scale $\Lambda_{QCD}$, which arises from dimensional transmutation, the Lagrangian of this theory contains the vacuum angle $\theta$ and the six quark mass parameters $m_u, m_d,m_s, m_c,m_b,m_t$. Their values are by now known quite accurately, but the observed pattern is bizarre and totally ungrasped. We also know that, in the basis where the quark mass matrix is diagonal, real and positive, the vacuum angle is tiny, but we do not know why that is so. 

To work out the properties of QCD is a fascinating challenge.
Models (AdS, CFT, superconductors, NJL \ldots) may help developing the intuitive understanding of QCD,
but they are a meagre replacement for the real thing.
Much work remains to be done to confront low energy precision experiments with our present understanding of the laws of nature: lattice simulations, effective field theory, dispersion theory, \ldots\, For an overview of ongoing work and references to the literature, I refer to \cite{FLAG}-\cite{Cirigliano}. 

\subsubsection*{Acknowledgement}
I thank Tian Yu Cao, Gerhard Ecker, Lev Okun, Zurab Silagadze and Misha Vysotsky for useful comments concerning the material presented here. Also, it is a pleasure to thank Nino Zichichi for a most enjoyable stay at Erice.

\end{document}